
\input phyzzx

\nopagenumbers

\Pubnum={CPTH-A213.1292\endline HUPD-9216}
\date={December 1992}

\def\mbox#1#2{\vcenter{\hrule \hbox{\vrule height#2in
                \kern#1in \vrule} \hrule}}  
\def\square{\,\raise.5pt\hbox{$\mbox{.09}{.09}$}\,}
\def\sqb{\,\raise.5pt\hbox{$\overline{\mbox{.09}{.09}}$}\,}

\titlepage
\title{Conformal dynamics of quantum gravity with torsion}

\author{I. Antoniadis}
\address{Centre de Physique Th\'eorique\break
Ecole Polytechnique\break
91128 Palaiseau, France\break}

\andauthor{S.D. Odintsov\foot{On leave from Pedagogical Institute, 634041
Tomsk, Russia}}
\address{Department of Physics\break
Hiroshima University\break
Higashi-Hiroshima 724, Japan}


\abstract{The trace anomaly induced dynamics of the conformal factor is
investigated in four-dimensional quantum gravity with torsion. The
constraints
for the coupling constants of torsion matter interaction are obtained in
the
infrared stable fixed point of the effective scalar theory.}

\endpage


\REF\Ant{I. Antoniadis and E. Mottola, Phys.\ Rev.\ {\bf D45} (1992) 2013.}
\REF\Odi{S.D. Odintsov. Z.\ Phys.\ {\bf C54} (1992), 531.}
\REF\Anto{I. Antoniadis, P.O. Mazur and E. Mottola, Conformal Symmetry and
 central charges in 4 dimensions, LA-UR 92 preprint.}
\REF\Rei{R.J. Reigert, Phys.\ Lett.\ {\bf B134} (1984) 56; E.T. Tomboulis,
 Nucl.\ Phys.\ {\bf B329} (1990) 410; S.D. Odintsov and I.L. Shapiro,
Class.\
 Quant.\ Grav.\ {\bf 8} (1991) L57.}
\REF\Pol{A.M. Polyakov, Mod.\ Phys.\ Lett.\ {\bf A2} (1987) 899; V.G.
Knizhnik,
 A.M. Polyakov and A.B. Zamolodchikov, Mod.\ Phys.\ Lett.\ {\bf A3} (1988)
819;
 J. Distler and H. Kawai, Nucl.\ Phys.\ {\bf B321} (1989) 509; F. David,
Mod.\
 Phys.\  Lett. {\bf A3} (1988) 1651.}
\REF\Leu{H. Leutwyler, Phys.\ Lett.\ {\bf 153} (1985) 65; L. Alvarez -
Gaum\'e
and P. Ginsparg, Ann.\ Phys.\ (NY) {\bf 161} (1985) 423; K. Li, Phys.\
Rev.\ {\bf
 D34} (1986) 2292; T. Fukuyama and K. Kamimura, Phys.\ Lett.\ {\bf B200}
(1988)
 75; S.D. Odintsov, Rivista\ Nuovo\ Cim.\ {\bf 15} (1992) No.3 .}
\REF\Buc{I.L. Buchbinder, S.D. Odintsov and I.L. Shapiro, Phys.\ Lett.\
{\bf 162}
(1985) 93; Rivista Nuovo\ Cim.\ {\bf 12} (1989) 1; see also Effective
action in
 quantum qravity, IOP Publishing, 1992 (Chapter 4,5).}
\REF\Bir{N. Birrell and P. Davies, Quantum fields in curved space, CUP,
1982 and
 references therein.}
\REF\Fra{E.S. Fradkin and A.A. Tseytlin, Nucl.\ Phys.\ {\bf B201} (1982)
469.}
\REF\Sez{E. Sezgin and P. van Nieuwenhuizen, Phys.\ Rev.\ {\bf D21} (1980)
 3269.}

\footline={\hss\tenrm\folio\hss}\pageno=1

In a recent paper [1] the trace anomaly induced dynamics of the
conformal factor in four-dimensional quantum gravity has been investigated.
The conformal sector was described by spacetime metrics of the form:
$$
g_{\mu\nu}(x) = e^{2\sigma(x)}\eta_{\mu\nu} \ ,
$$
where $e^{\sigma}$ is the conformal factor and $\eta_{\mu\nu}$ is the flat
metric. The generalization to curved fiducial metric was considered in
refs.[2,3]. The effective action for $\sigma$ can be obtained by
integrating
the general form of trace anomaly in curved spacetime [4,3]. It turns out
that
the resulting action is local in $\sigma$ and possesses a non-trivial,
infrared
stable critical point which can be computed to all orders in perturbation
theory [1]. In particular, the anomalous scaling dimension of the conformal
field $e^\sigma$ was found using the same line of reasoning as in
two-dimensional gravity [5]. It was argued that the effective $\sigma$
theory
describes four-dimensional quantum gravity at large distances and provides
a
framework for a dynamical solution of the cosmological constant problem
[\Ant].

The purpose of this note is to generalize the results of ref.[1] in the
case of
quantum gravity with torsion. Torsion interactions start becoming
non-trivial in
four dimensions. In fact in two dimensions there is no minimal coupling of
matter with torsion. Furthermore, in the non-minimal case [6] torsion
appears
quadratically and, after functional integration, 2d induced gravity with
torsion
is equivalent to 2d gravity without torsion up to coupling constants
redefinitions. In four dimensions minimal interactions of spinors with
torsion
are non-trivial. Moreover, the torsion dependence of trace anomaly is much
more
complicated than in two dimensions [7], so that one may expect to find some
new
properties of the conformal factor dynamics in this case.

Let us start with the free actions for scalar and spinor fields in curved
spacetime in the presence of torsion $T^\alpha_{\gamma\beta}$. Recall that
in a
theory with torsion the connection $\Gamma^\alpha_{\beta\gamma}$ is not
symmetric in the lower indices:
$$
   \Gamma^\alpha_{\beta\gamma}-\Gamma^\alpha_{\gamma\beta}
        =T^\alpha_{\gamma\beta} \neq0.
\eqno(1)
$$
This relation is the origin of torsion. Here, we consider the case where
the
only non-vanishing components of torsion are those which correspond to the
totally antisymmetric part, $S^\mu\equiv
i\varepsilon^{\alpha\beta\nu\mu}T_{\alpha\beta\nu}$, called pseudotrace.
This
is the only part of torsion which can interact minimally with matter (see
ref.[7] for more details).

The conformally invariant quadratic actions for scalars ($S_0$) and spinors
($S_{1 \over 2}$) are [7]:
$$
S_0= {1 \over 2}\int d^4x{\sqrt
{-g}}\{g^{\alpha\beta}\partial_\alpha\varphi
\partial_\beta\varphi+{1 \over 6}R\varphi^2+\xi_0 S_\mu S^\mu\varphi^2\},
\eqno(2)
$$
$$
S_{1 \over 2}=i\int d^4x{\sqrt {-g}}\{{\bar \psi}
(\gamma^\mu\nabla_\mu-\xi_{1\over 2}
               \gamma_5\gamma^\mu S_\mu)\psi\},
\eqno(3)
$$
where $\xi_0,\xi_{1\over 2}$ are arbitrary constants. These actions are
invariant
under the conformal transformations:
$$
        g_{\mu\nu}\rightarrow e^{2\sigma}g_{\mu\nu},\hskip 10pt
           \varphi\rightarrow e^{-\sigma}\varphi,\hskip 10pt
              \psi\rightarrow e^{-{{3\sigma} \over 2}}\psi
\eqno(4)
$$
for any values of $\xi_0$, $\xi_{1\over 2}$. The choice $\xi_0=0$,
$\xi_{1\over
2}={1 \over 8}$ corresponds to the minimal interaction of matter with
torsion.
Here we do not consider torsion interaction with gauge fields. Such
interaction
is non minimal, and it exists in general only in the abelian case leading
to
terms which are first order in quantum fields [7]. It is therefore
irrelevant
for our purposes.

The general form of the trace anomaly for free conformally invariant matter
in curved spacetime with torsion is:
$$
\eqalign{
T_\mu^\mu = &b (F+{2 \over 3}\square R)+b' G+b^{''}\square R
        +a_1F_{\mu\nu}F^{\mu\nu} \cr
        &+ a_2(S_\mu S^\mu)^2+a_3\square(S_\mu S^\mu)+a_4\nabla_\mu
        (S_{\nu}\nabla^\nu S^{\mu}-S^{\mu}\nabla_\nu S^{\nu}), }
\eqno(5)
$$
where
$$
F\equiv C_{\mu\nu\alpha\beta}C^{\mu\nu\alpha\beta}, \ \
G=R^2_{\mu\nu\alpha\beta}-4R^2_{\mu\nu}+R^2, \ \
F_{\mu\nu}=\nabla_\mu S_\nu-\nabla_\nu S_\mu,
$$
$$\eqalign{
&b={1 \over {120(4\pi)^2}}(N_0+6N_{1 \over 2}+12N_1), \ \
b'=-{1 \over {360(4\pi)^2}}(N_0+11N_{1 \over 2}+62N_1), \cr
&a_1=-{2 \over {3(4\pi)^2}}\sum\xi_{1\over 2}^2, \ \
a_2={1 \over {2(4\pi)^2}}\sum\xi_0^2, \ \
a_3={1 \over {3(4\pi)^2}}\sum(2\xi_{1\over 2}^2-{1 \over 2}\xi_0^2), \cr
&a_4=-{2 \over {3(4\pi)^2}}\sum\xi_{1\over 2}^2\ , }
$$
with $N_0$, $N_{1 \over 2}$ and $N_1$ the number of scalars, Dirac fermions
and
vectors. The coefficients $b$ and $b'$ can be found in ref.[8], $b^{''}$ is
arbitrary since it can be changed by adding a local $R^2$ term in the
action,
while the coefficients $a_1$, $a_2$, $a_3$ and $a_4$ are given in ref.[7]
for
the case of minimal interaction with torsion ($\xi_0=0$, $\xi_{1\over 2}={1
\over 8}$)  $a_2=0$.

We now choose the conformal parametrization $g_{\mu\nu}=e^{2\sigma(x)}{\bar
g}_{\mu\nu}$ and $S_\mu={\bar S}_\mu$, where ${\bar g}_{\mu\nu}$ and ${\bar
S}_\mu$ are the fiducial metric and torsion, respectively. The conformal
factor
acquires dynamics by an effective action whose $\sigma$-variation
reproduces the
anomalous trace (5):
$$
\eqalign{
T^\mu_\mu=&bF+b'(G-{2 \over 3}\square R)+[b^{''}+{2 \over 3}(b+b^{''})]\square
R
+a_1F_{\mu\nu}F^{\mu\nu}\cr + &a_2(S_\mu S^\mu)^2+a_3\square(S_\mu S^\mu)
+a_4\nabla_\mu(S_\nu \nabla^\nu S^\mu-S^\mu\nabla_\nu S^\nu) \cr
{\equiv}&{1 \over {\sqrt{-g}}} {\delta \over {\delta\sigma(x)}}
S_{anom}(\sigma)}
 \eqno(6)
$$
Integrating eq. (6) we get (see refs.[1,5] for $S_\mu=0$, and ref.[7] for
$S_\mu\neq 0$) :
$$
\eqalign{
S_{anom}[\sigma]= & \int d^4 x \sqrt{-{\bar g}}\{\sigma[b{\bar F} +b'({\bar
G}
    -{2 \over 3}{\sqb}{\bar R})+a_1{\bar F}^2_{\mu\nu}+a_2{\bar S}^4]
 \cr
    &+2b'\sigma{\bar \Delta}\sigma-a_3{\bar \nabla}_\mu \sigma{\bar
\nabla}^\mu {\bar S}^2-a_4{\bar \nabla}_\mu \sigma({\bar S}_\nu {\bar
\nabla}^\nu {\bar S}^\mu -{\bar S}^\mu {\bar \nabla}_\nu {\bar S}^\nu ) \cr
    &+(a_3+{1 \over 2}a_4 ){\bar S}^2{\bar \nabla}_\mu \sigma\nabla^\mu
\sigma +
       a_4 {\bar S}^\mu{\bar S}^\nu{\bar \nabla}_\mu \sigma{\bar
\nabla}_\nu
      \sigma\cr
    &-{1 \over 12}[b^{''}+{2 \over 3}(b+b')]\int d^4 x\sqrt{-{\bar
g}}[{\bar R}
     -6{\sqb}\sigma-6({\bar \nabla}_\mu \sigma)({\bar \nabla}^\mu
      \sigma)]^2\}, \cr}
\eqno(7)
$$
where ${\bar S}^2={\bar S}_\mu{\bar S}^\mu$, and $\Delta$ is the
conformally
covariant fourth-order operator acting on scalars,
$$
\Delta=\square^2+2R^{\mu\nu} \nabla_\mu \nabla_\nu-{2 \over
3} R\square+{1 \over 3}\nabla_\mu R \nabla^\mu.
\eqno (8)
$$
In eq. (7) we have omitted $\sigma$-independent terms.

Now as in two dimensions [5] we add $S_{anom}$ to the classical gravity
action:
$$\eqalign{
S_{cl}=&{1 \over {2\kappa}}\int d^4 x \sqrt{-g}(R+hS_\mu S^\mu-2\Lambda)\cr
=&{1 \over {2\kappa}}\int d^4 x\sqrt{-{\bar g}}\{e^{2\sigma}
[R+6({\bar\nabla}_\mu \sigma)({\bar\nabla}^\mu \sigma)]-2\Lambda
 e^{4\sigma}+h e^{2\sigma}{\bar S}^2\},\cr }
\eqno (9)
$$
where $h$ is a constant parameter (e.g. for the Einstein-Cartan theory
$h=-{1
\over 24}$). Thus, the total effective action is:
$$
S_{eff}=S_{anom}+S_{cl}                                              \eqno
(10)
$$

When the conformal factor is quantized, the requirement of general
covariance
in the vacuum state of $\sigma$-theory implies the vanishing of the total
trace at the quantum level [\Anto]. Therefore, the $\beta$-functions of
all couplings appearing in the effective $\sigma$-action must vanish.
Without
loss of generality, let us consider the effective theory (10) with
flat fiducial metric and arbitrary constant torsion background $S_\mu$:
$$
\eqalign{
S_{eff}=&-{{Q^2} \over {(4\pi)^2}}\int d^4 x (\square \sigma)^2-\zeta\int
d^4
 x[2\alpha(\partial_\mu \sigma)^2\square \sigma +\alpha^2(\partial_\mu
 \sigma)^4]\cr
&+\gamma \int d^4 xe^{2\alpha \sigma}(\partial_\mu \sigma)^2-{\lambda \over
 {\alpha^2}}\int d^4 xe^{4\alpha \sigma}+\int d^4 x \{(a_3+{1 \over
 2}a_4)S^2(\partial_\mu \sigma)^2 \cr
&+a_4S^\mu S^\nu \partial_\mu \sigma \partial_\nu \sigma +{h \over
 {2\kappa\alpha^2}}e^{2\alpha \sigma}S^2 + a_2\sigma S^4\}, }
    \eqno (11)
$$
where the transformations $\sigma\rightarrow \sigma \alpha$,
 $S_{eff}\rightarrow\alpha^{-2} S_{eff}$ are made, and the notations of
ref.[1]
are introduced,
$$
{{Q^2} \over {(4\pi)^2}}=2b+3b',\ \   \zeta=2b+2b'+3b^{''}, \ \
\gamma ={3\over \kappa}, \ \ \lambda ={\Lambda \over \kappa}\ .
\eqno (12)
$$
$\alpha$ is the anomalous scaling dimension of the conformal factor at the
non-trivial infrared critical point we want to determine.

A simple power counting argument shows that the effective action (11) is
renormalizable in $\sigma$-perturbation theory if we add the action of the
external torsion-field:
$$
S_{ext}={\eta \over {\alpha^2}}\int d^4 x(S_\mu S^\mu)^2.
                                                                  \eqno
(13)
$$
The $\zeta$ coupling renormalization does not depend on the remaining
interactions of (11), and a straight-forward one-loop calculation of its
$\beta$-function shows that $\zeta =0$ is an infrared stable fixed point
[1].
At $\zeta =0$, the contribution to the anomaly coefficients $b$ and $b'$ of
the
$\sigma$-field itself is coming from the quartic operator $\Delta$ (8) and was
computed in ref.[\Anto]. Moreover, due to the quartic kinetic operator, the
effective action (11) becomes superrenormalizable and all $\beta$-functions
can
be computed exactly. The presence of torsion background does not modify the
$\beta$-functions of $\gamma$ and $\lambda$ couplings. Their vanishing
imply [1]:
$$
\alpha_\pm={{1\pm \sqrt{1-4/Q^2}} \over {2/Q^2}} ,\ \ \ \ {\lambda \over
 \gamma^2}={{2\pi^2} \over {Q^2}}[1+{{4\alpha^2} \over
 {Q^2}}+{{6\alpha^4} \over {Q^4}}].               \eqno (15)
$$
Eq. (15) determines at the fixed point the anomalous scaling dimension
$\alpha$
in terms of the ``central charge" $Q^2$, and relates the cosmological with
the
Newtonian couplings. Now we can calculate the additional $\beta$-functions
connected with the torsion.

We first perform a one-loop computation using the standard algorithm [9]
and
then we will show that the one-loop result is exact. Expanding $\sigma$
around
a classical background solution and keeping only terms which are quadratic
in
quantum fluctuations, we can write the resulting kinetic operator
corresponding
to the action (11) in the following form:
$$
\hat {H}=\square^2+D^{\mu\nu}\nabla_\mu \nabla_\nu+H^\mu\nabla_\mu+p
                                                                 \eqno (16)
$$
where $H^\mu$ does not give any contribution to the one-loop divergences
and can
 be dropped,
$$
D^{\mu\nu}={{(4\pi)^2} \over {Q^2}}[\gamma
e^{2\alpha\sigma}g^{\mu\nu}+(a_3+{1
 \over 2}a_4)S^2g^{\mu\nu}+a_4S^\mu S^\nu],
$$
$$
p=-{{(4\pi)^2} \over {Q^2\kappa}} he^{2\alpha\sigma}S^2-2({{4\pi} \over
 Q})^2\gamma\alpha^2e^{2\alpha\sigma}(\partial_\mu
\sigma)^2+8\lambda({{4\pi}
 \over Q})^2 e^{4\alpha\sigma}\ .
$$
The corresponding $a_2$ Schwinger-De Witt coefficient in the heat kernel
expansion of $\hat {H}$ is:
$$
\eqalign{
a_2(\hat{H})&={1 \over {(4\pi)^2}}\{-p+{1 \over 48}D^2+{1 \over
 24}D_{\mu\nu}D^{\mu\nu}\}
                  \cr
&=e^{2\alpha\sigma}S^2[{h \over {\kappa Q^2}}+{{\gamma(4\pi)^2} \over
 {Q^4}}(a_3+{3\over 4}a_4)]+S^4{{(4\pi)^2} \over {Q^4}}[{1 \over 2}a^2_3+{3
 \over 4}a_3a_4+{5 \over 16}a^2_4], }
                                                                     \eqno
(17)
$$
where we neglected total derivatives, like $\square D$ and $\partial_\mu
\partial_\nu D^{\mu\nu}$, as well as $S$ independent terms which have been
already discussed. Taking also into account the ``classical" contribution
due to
the change of the scaling dimension of the conformal factor $e^\sigma$ we
get:
$$
\eqalign{
& \beta_h=(2-2\alpha+{{2\alpha^2} \over {Q^2}}){h \over
 {2\kappa}}+{{\alpha^2\gamma(4\pi)^2} \over {Q^4}}(a_3+{3 \over 4}a_4), \cr
& \beta_{\eta}={{\alpha^2(4\pi)^2} \over {Q^4}}[{1 \over 2}a^2_3+{3 \over
4}a_3
 a_4+{5 \over {16}}a^2_4]. }
 \eqno (18)
$$

One can now show by a Feynman graph analysis similar to the one done in
ref.[1]
that $\beta_h$ is the exact $\beta$-function and there are no higher order
corrections. In fact, all additional terms in (11) involving the torsion
background amount only to a change of $\sigma$-propagator, with the
exception
of the $h$-term which generates new vertices. A simple power counting
argument
then shows that the only graphs with primitive divergences must contain at
most
two $\gamma$-vertices or exactly one $\lambda$ or $h$-vertex. It follows
that
the $S$-dependent part of divergences come from tadpole diagrams with one
$h$
or $\gamma$-vertex and lead to the expression for $\beta_h$ in (18).
$\beta_{\eta}$ is associated to vacuum ($\sigma$-independent) terms which
can be
made zero by an appropriate $\sigma$-translation, due to the presence of
the
last linear term in $\sigma$ in (11).

The gravitational (spin-two and reparametrization ghost) contribution
remains
an open question. In ref.[3], it was argued that in the infrared this
amounts
only to a change of the trace-anomaly coefficients (5) in the effective
$\sigma$-theory. Futhermore, this contribution was calculated at the
one-loop
level using the Einstein action, as well as the conformally invariant but
four-derivative Weyl tensor squared action. In the presence of non-vanishing
torsion the computation becomes more complicated. A potentially interesting
case
could be the example of higher derivative gravity with torsion which seems
to
avoid the usual unitarity problem for some values of the parameters at the
classical level [10].

Focussing in the $\sigma$-field sector, and using eq.
(15), the vanishing of $\beta_h$ gives
$$
a_3 + {3 \over 4}a_4 = 0.
\eqno (19)
$$
This condition constraints the torsion parameters of the theory at the
infrared stable fixed point. In the case of pure matter it implies
$\xi_0=\xi_{1\over 2}$, which for minimal interaction is satisfied only for
zero torsion.

\bigskip
\ack{This work was supported in part by the EEC contracts SC1-0394-C and
SC1-915053. SDO wishes to thank Particle Theory Group at Hiroshima
University for
kind hospitality and JSPS for financial support. We also thank I.L. Shapiro
for
discussions on the point that $\beta_h$ is given by the one-loop result.}

\refout
\bye